\documentclass{book}    % fundamental class file is 'book'.
\usepackage{meta}  % use the file 'meta.sty'.
\usepackage{graphics} 
\begin{document}
 \title{Sub-wavelength diffraction-free  imaging with low-loss metal-dielectric multilayers }
\maketitle

\author      {R. Kotynski}
\affiliation {University of Warsaw}
\address     {}% optional
\city        {Warsaw}
\postalcode  {}% optional
\country     {Poland}
\phone       {}    % optional
\fax         {}    % optional
\email       {}  % optional
\misc        {}  % optional
\nomakeauthor
%------------------------------------

%=== List of authors (in order) ========
%-- Author(s) for the second affiliation ---
\author      {T. Stefaniuk}
\affiliation {University of Warsaw}
\address     {}% optional
\city        {Warsaw}
\postalcode  {}% optional
\country     {Poland}
\phone       {}    % optional
\fax         {}    % optional
\email       {}  % optional
\misc        {}  % optional

\author      {A. Pastuszczak}
\affiliation {University of Warsaw}
\address     {}% optional
\city        {Warsaw}
\postalcode  {}% optional
\country     {Poland}
\phone       {}    % optional
\fax         {}    % optional
\email       {}  % optional
\misc        {}  % optional

\nomakeauthor

 \begin{authors}
 {\bf R. Kotynski}, {\bf T. Stefaniuk}, {\bf A. Pastuszczak}\\
 \medskip
 University of Warsaw, Faculty of Physics, Poland\\
 rafal.kotynski@fuw.edu.pl, tstefaniuk@igf.fuw.edu.pl, apastuszczak@igf.fuw.edu.pl
 \end{authors}

\begin{paper}

\begin{metaabstract}
We demonstrate numerically the diffraction-free propagation of sub-wavelength sized optical beams through simple elements built of metal-dielectric multilayers. The proposed metamaterial consists of silver and a high refractive index dielectric, and is  designed using the effective medium theory as strongly anisotropic and impedance matched to air. Further it is characterised with the transfer matrix method, and investigated with FDTD. The diffraction-free behaviour is verified by the analysis of FWHM of PSF in the function of the number of periods. Small reflections, small attenuation, and reduced Fabry Perot resonances make it a flexible diffraction-free material for arbitrarily shaped optical planar elements with sizes of the order of one wavelength. 
\end{metaabstract}

%---Content of Paper Text-----------------------
\psection{Introduction}
%\section{Introduction}
Since the seminal paper by Pendry~\cite{pendry2000nrm}, subwavelength imaging at visible wavelengths has been demonstrated in much
thicker low-loss layered silver-dielectric periodic structures~\cite{wood2006,Scalora2007opex,Scalora08pra,Scalora:OE09,belov2005csi,belov2006prb,Li:prb07,Kotynski:jopa2009}. In fact, the effective medium theory (EMT)  provides sufficient means to explain enhanced transmission through the metal-dielectric stacks. Subwavelength resolution results from the extreme anisotropy of the effective permittivity tensor~\cite{wood2006,belov2005csi,belov2006prb}. 

Let us continue the introduction by linking the metal-dielectric multilayers for sub-wavelength imaging with the concepts taken from Fourier Optics.
 We refer to the model of a linear shift invariant scalar system (LSI)~\cite{Saleh,GoodmanFourierOptics} for the description of optical multilayers in a situation when they act as imaging nano-elements for coherent monochromatic light. In-plane imaging through a layered structure consisting of uniform and isotropic materials represents a LSI, for either TE or TM polarisations. Linearity of the system is the consequence of linearity of materials and validity of the superposition principle for  optical fields. Shift invariance results from the assumed infinite perpendicular size of the multilayer and the freedom in the choice of an optical axis. 

A scalar description is valid for the TE and TM polarisations in 2D since all other field components may be derived from $E_y$ or $H_y$, respectively, where the co-ordinate system is oriented as in fig.~\ref{fig.schem}.

For the TM polarization, the magnetic field $H_y(x,z)$ may be represented with its spatial spectrum $\hat H_y(k_x,z)$
\begin{equation}
 H_y(x,z)=\int_{-\infty}^{+\infty} \hat H_y(k_x,z) exp(\imath k_x x) dk_x,
\end{equation} 
where, at least for lossless materials, the spatial spectrum is clearly separated into the propagating part $k_x^2<k_0^2 \epsilon$ and evanescent part $k_x^2>k_0^2 \epsilon$.
\begin{figure}
a.\includegraphics[width=5cm]{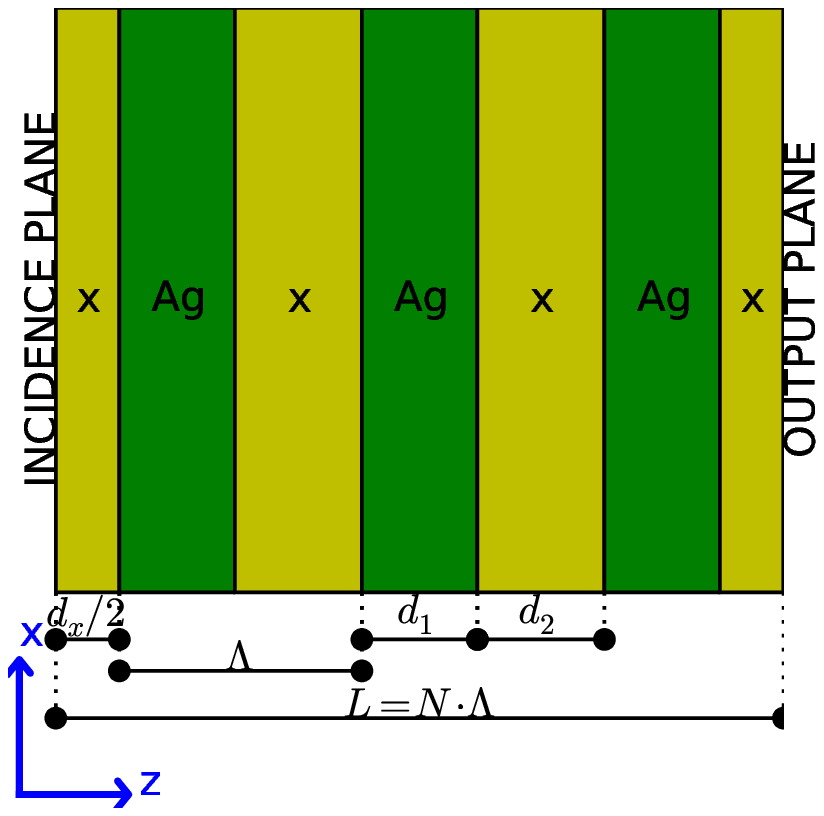}
b.\includegraphics[width=5cm]{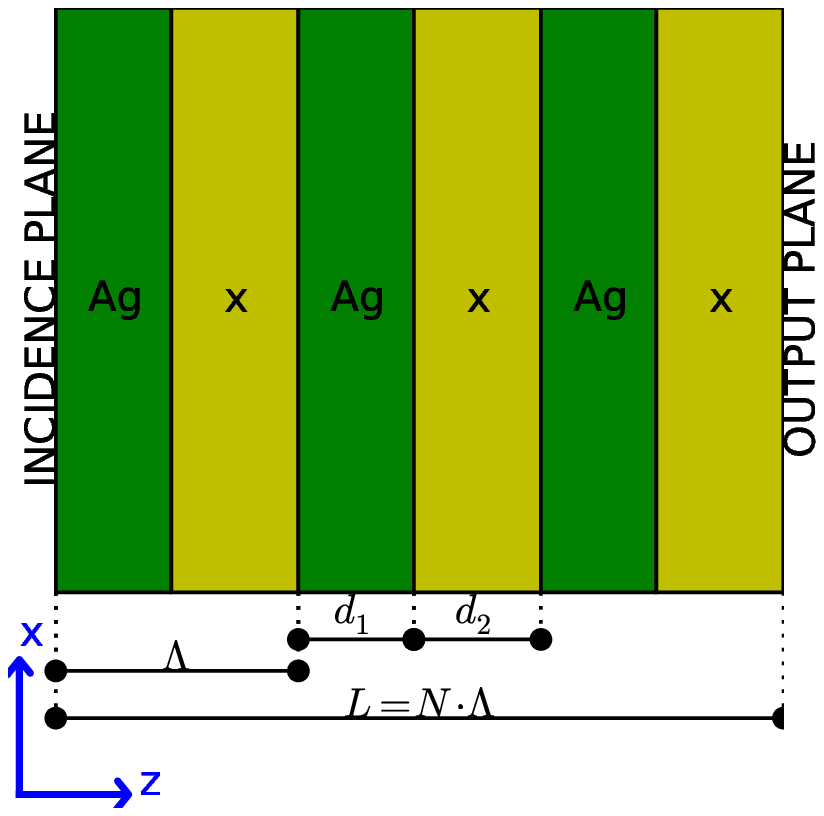}
c.\includegraphics[width=5cm]{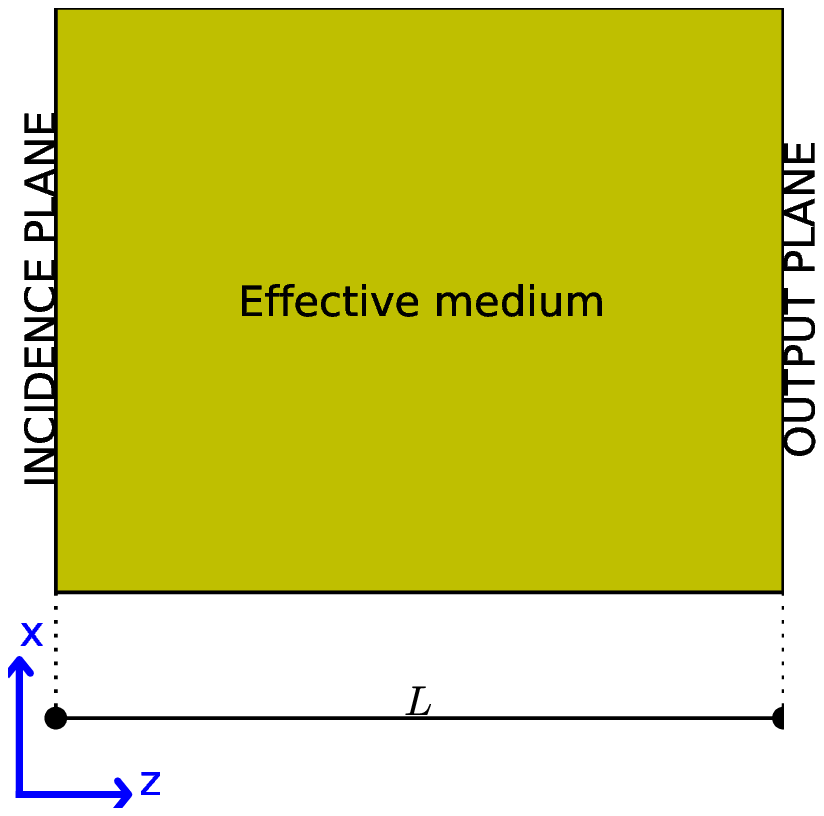}
\caption{Schematic of a periodic silver-dielectric multilayer with  symmetric (\textbf{a})  or  non-symmetric (\textbf{b}) composition, and the equivalent effective medium slab (\textbf{c}).
\label{fig.schem}}
\end{figure} 
The transfer function $t(k_x)$ (TF) is the ratio of the  output to incident fields spatial spectra   and corresponds to the amplitude transmission coefficient of the multilayer
\begin{equation}
\hat H_y(k_x,z=L)=t(k_x) \cdot \hat H_y^{Inc}(k_x,z=0).
\label{eq.mtfdef}
\end{equation}
Due to reflections, the incident field $\hat H_y^{Inc}(k_x,z=0)$ differs from the total field $\hat H_y(k_x,z=0)$.

The point spread function (PSF) is the inverse Fourier transform of the TF and has the interpretation of the response of the system to a point signal $\delta(x)$. The response to an arbitrary input $H_y^{Inc}(x,z=0)$ can be further expressed as its convolution with the PSF
\begin{equation}
 H_y(x,z=L)=H_y^{Inc}(x,z=0)* PSF(x)
\label{eq.psfconool}
\end{equation}

 PSF of an imaging system usually provides clear information about the resolution, loss or enhancement of contrast, as well as the characteristics of image distortions. However, for subwavelength imaging, the PSF is not a straightforward measure of resolution and even imaging of objects smaller than the FWHM of PSF is possible~\cite{Kotynski:arXiv-09}.

In this paper we demonstrate  a diffraction-free material for subwavelength sized optical beams. We  combine the following properties of the multilayer:  PSF with sub-wavelength size and little dependence on the thickness of the structure, high transmission, low losses, and a limited dependence of the imaging properties on the size of external layers. Together, these properties allow to use a multilayer as a flexible construction material for various optical imaging nano-devices.

\begin{figure}
a.\includegraphics[width=7.5cm]{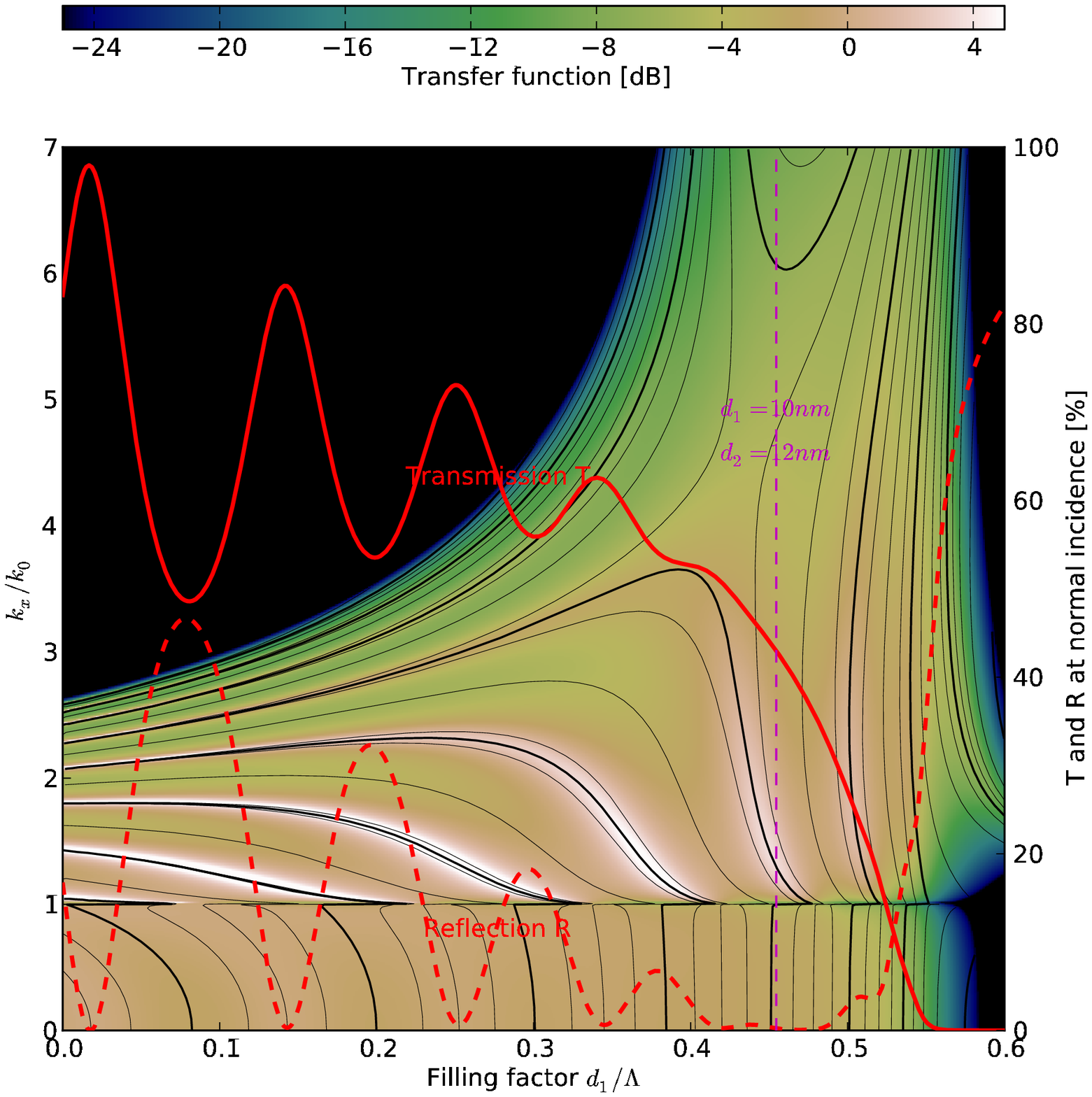}
b.\includegraphics[width=8.5cm]{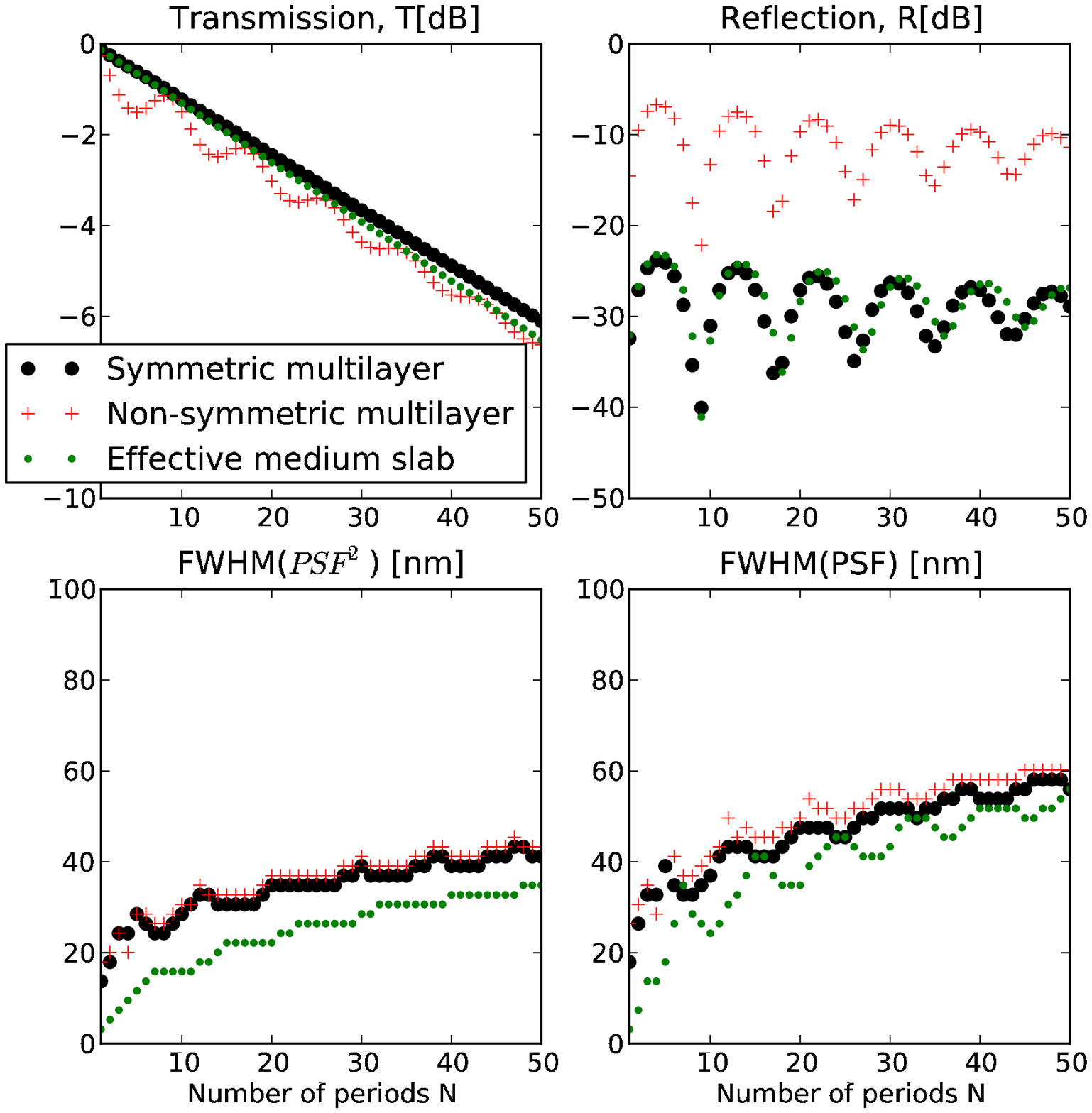}
\caption{\textbf{a} Intensity transmission and reflection coefficients ($T,R$) of the symmetric multilayer as a function of the filling factor, plotted over the transfer function (in vertical cross-sections of the color map). Phase isolines are separated by $\pi/4$. The pitch $\Lambda=d_1+d_2$ is fixed at $22nm$, and the total thickness is $L=30\Lambda$. \textbf{b.}~Characteristics of the symmetric and non-symmetric multilayers, and of the equivalent slab made of an effective medium, in the function of the number of periods. Intensity transmission and reflection coefficients calculated at normal incidence (top); FWHM of $PSF$ and $\arrowvert PSF\arrowvert^2$ (bottom).\label{fig.tf}}
\end{figure} 

\psection{Imaging with sub-wavelength resolution in metal-dielectric multilayers} 
The dispersion relation of a two-component infinite stack for the TM polarisation has the form~\cite{Wu:PRB67-235103}

\begin{equation}
\cos(K_z \Lambda)=\cos(k_{1z} f \Lambda) \cos(k_{2z} (1-f) \Lambda)
 -{ \frac{\sin(k_{1z} f \Lambda) \sin(k_{2z} (1-f) \Lambda)}{2}}\left({\frac{k_{1z} \epsilon_2}{\epsilon_1 k_{2z}}} + {\frac{k_{2z} \epsilon_1}{\epsilon_2 k_{1z}}}\right)\label{eq.dispinf},
\end{equation}

where $K_z$ is the Bloch wavenumber, $\Lambda=d_1+d_2$ is the period of the stack, $d_i$ and $\epsilon_i$ are the layer thickness and permittivity of material $i=1,2$, the filing fraction of material $1$ is $f=d_1/\Lambda$ and the local dispersion relations are $k_{iz}^2+k_x^2=k_0^2 \epsilon_i$. Wavevector $k_x$ is conserved at the layer boundaries and depends on the incidence conditions. The Bloch wavenumber $K_z$ is real for a Bloch mode in a lossless stack, however for evanescent waves in a finite stack or for lossy materials $K_z$ may be complex.
In the first BZ, the real part of $K_z$  satisfies $\pi/\Lambda<=re(K_z)<=\pi/\Lambda$. The group velocity as well as the imaginary part responsible for absorption do not depend on the choice of BZ. 

When the layers are thin $K_z \Lambda, k_{iz} \Lambda<<1$, the second order expansion of (\ref{eq.dispinf}) over the arguments of trigonometric functions leads to the dispersion relation for an uniaxially anisotropic effective medium
\begin{equation}
K_z^2/\epsilon_x+k_x^2/\epsilon_z=k_0^2,
\end{equation}
with $\epsilon_x = f \epsilon_1  + (1-f) \epsilon_2$ and $\epsilon_z =1/( f /\epsilon_1  + (1-f)/ \epsilon_2)$. This is the basis of the  effective medium approximation thoroughly discussed by Wood et al.~\cite{wood2006}, also as the basis for applying the near-field approximation. The transmission coefficient of the Fabry-Perot (FP) slab consisting of the homogenized effective medium is then given as~\cite{wood2006},
\begin{equation}
t(k_x)=\left( \cos(k_z L)-0.5\imath (K_z\epsilon_x /k_z\epsilon+k_z\epsilon /K_z\epsilon_x )\sin(k_z L)) \right)^{-1},
\end{equation}
where $k_z$ and $\epsilon$ refer to the external medium, and $L$ is the total thickness of the slab. At the same time $t(k_x)$ is the already mentioned coherent amplitude transfer function of the imaging system~\cite{GoodmanFourierOptics}. When $\lvert \epsilon_x/\epsilon_z\rvert<<1$ (which is satisfied for a lossless metal for $\epsilon_1/\epsilon_2=-d_1/d_2$~\cite{belov2006prb}), for certain slab thickness the resonant FP condition becomes independent on the angle of incidence. This happens when $L=\lambda m/\sqrt{\epsilon_x}, m\in N$. Then $t(k_x)\equiv exp(- \imath K_z L)$ and the FP slab introduces the same phase shift for all harmonics of the spatial spectrum. Belov and Hao~\cite{belov2006prb} proposed to combine this condition with impedance matching between the external medium and the effective FP slab $\epsilon=\epsilon_x=f \epsilon_1  + (1-f) \epsilon_2$ and referred to that regime as canalization. However,  Li et al.~\cite{Li:prb07} questioned the importance of impedance matching in favour of the FP resonance condition. In fact, for a lossless metal and dielectric, the FP resonance is sufficient to entirely eliminate reflections resulting in perfect imaging $t(k_x)\equiv 1$ without impedance matching. 
In reality, losses make the condition $\epsilon=\epsilon_x$ only approximate and, at the same time, the finite value of $\Lambda$ limits the validity of homogenisation. Moreover, transmission through a finite slab strongly depends on the material and thickness of the external layers and appears to be the largest for a symmetrically designed slab~(Fig.~\ref{fig.schem}a) with half-width dielectric layers located at the boundaries~\cite{Scalora2007opex}. 
\psection{Numerical demonstration of diffraction-free propagation of sub-wavelength sized optical beams}
 After recalling the theory of sub-wavelength imaging in metal-dielectric multilayers, we  now focus on an imaging regime which may be called diffraction-free. We note, that the FP resonances are accompanied with a field pattern inside the slab similar to a standing wave~\cite{wood2006,Kotynski:jopa2009}. Therefore look for structures for which $t(k_x)\approx const$ nonetheless FP resonances are weak. 
 
Let us now focus on an example of a metal-dielectric periodic multilayer consisting of silver and high refractive index dielectric, which enables us to demonstrate and explain the diffraction-free propagation of sub-wavelength sized optical beams.   The structure operates at the wavelength of $\lambda=422nm$, when the permittivity of silver is equal to $\epsilon_1=-5.637 + 0.214\imath$\cite{JohnsonChristy}, and the permittivity of the dielectric such as $TiO_2$ or $SrTiO_3$~\cite{Saleh,Palik} is $\epsilon_2=(2.6)^2$. Layer thicknesses are assumed to be $d_1=10nm$, $d_2=12nm$ with periodic  symmetric or non-symmetric composition shown in fig.~\ref{fig.schem}a,b. The fraction $d_1/d_2$ may be  justified with the use of EMT. The corresponding permittivity of the effective medium gives $\epsilon_x\approx1.02+0.1\imath$ and $\epsilon_z\approx-158+191\imath$, which assure impedance matching with air $\epsilon=1\approx\epsilon_x$ together with the condition for the extreme anisotropy $\lvert \epsilon_x/\epsilon_z\rvert<<1$. The imaginary part of the refractive index $n_x=\sqrt{\epsilon_x}$ is reduced by the factor of $50$ compared to bulk silver, resulting in low-loss transmission. In fig.\ref{fig.tf}a we show the transfer function calculated rigorously with TMM for a range of filling factors $0<d_1/\Lambda<0.6$, when the total thickness of the structure is fixed at $L=660nm$. Indeed, the impedance matching which occures when $d_1=10nm, d_2=12nm$ results in reduced reflections, high transmission of both propagating and evanescent spatial harmonics, and a flat phase of the transfer function for a broad range of $k_x/k_0$. The size of the corresponding PSF is of the order of $\lambda/10\approx 40nm$ assuring the super-resolving properties of the device. Fig.~\ref{fig.tf}b shows the FWHM of PSF 
alongside with the reflection and transmission coefficient in the function of number of periods, for the symmetric, non-symmetric and effective-index composition of the structure. The major properties of the structure are the following: the size of PSF varies slowly with the size of the structure and is almost the same for the symmetric and non-symmetric composition, the FP resonances observed in transmission are weak (as opposed to those observed in reflections), the attenuation is uniform as the number of layers is increased. These properties assure an almost diffraction-free propagation through the structure, with a similar attenuation and the size of PSF for symmetric and non-symmetric composition and for a broad range of structure sizes $L$. This behaviour is illustrated in fig.~\ref{fig.fdtd1}a with an FDTD~\cite{Farjadpour2006ol} simulation showing the uniform non-diverging distribution of the Poynting vector inside the structure for a beam size of the order of $\lambda/10$. Transmission takes place along the direction normal to the layer boundaries with negligible divergence. Furthermore, our material may be used to fabricate slabs cut at arbitrary angle to layer surfaces. As is shown in fig.~\ref{fig.fdtd1}b,c we continue to observe the diffraction-free propagation for inclined layers, whether the aperture covers only a single or multiple periods of the grating. This may be heuristically explained with the similar PSF for a symmetric and non-symmetric composition of multilayers. Both of them are encountered at the cross sections drawn along the propagating beam, starting from various points inside the aperture and normal to the layer boundaries. Furthermore, it is possible to construct other simple optical nano-elements. In fig.~\ref{fig.fdtd2} we demonstrate the operation of a double multilayer, which may be part of a cloaking device or an optical interconnect, as well as prism for imaging subwavelength beams.

\begin{figure}
a.\includegraphics[height=4.5cm]{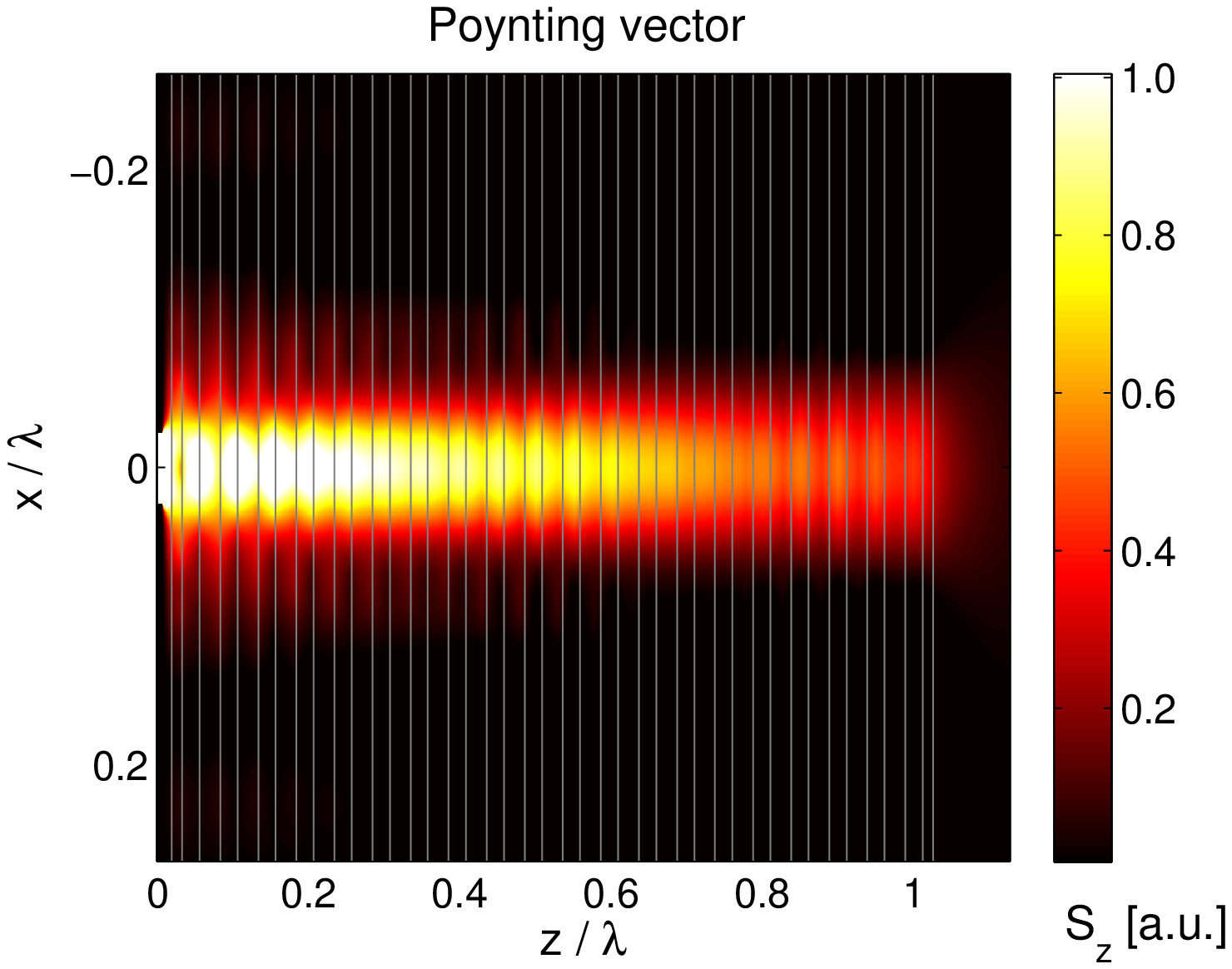}
b.\includegraphics[height=5cm]{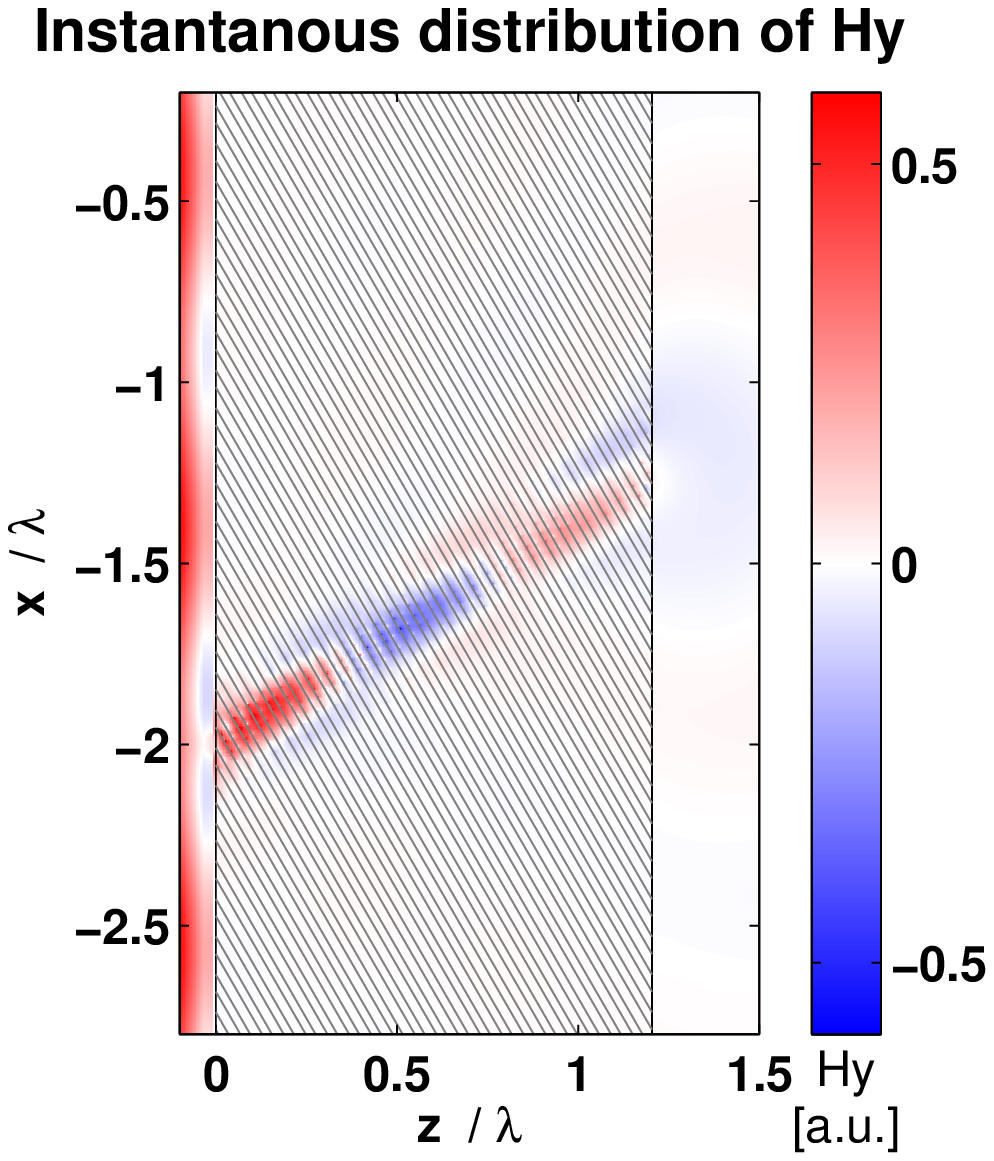}
c.\includegraphics[height=5cm]{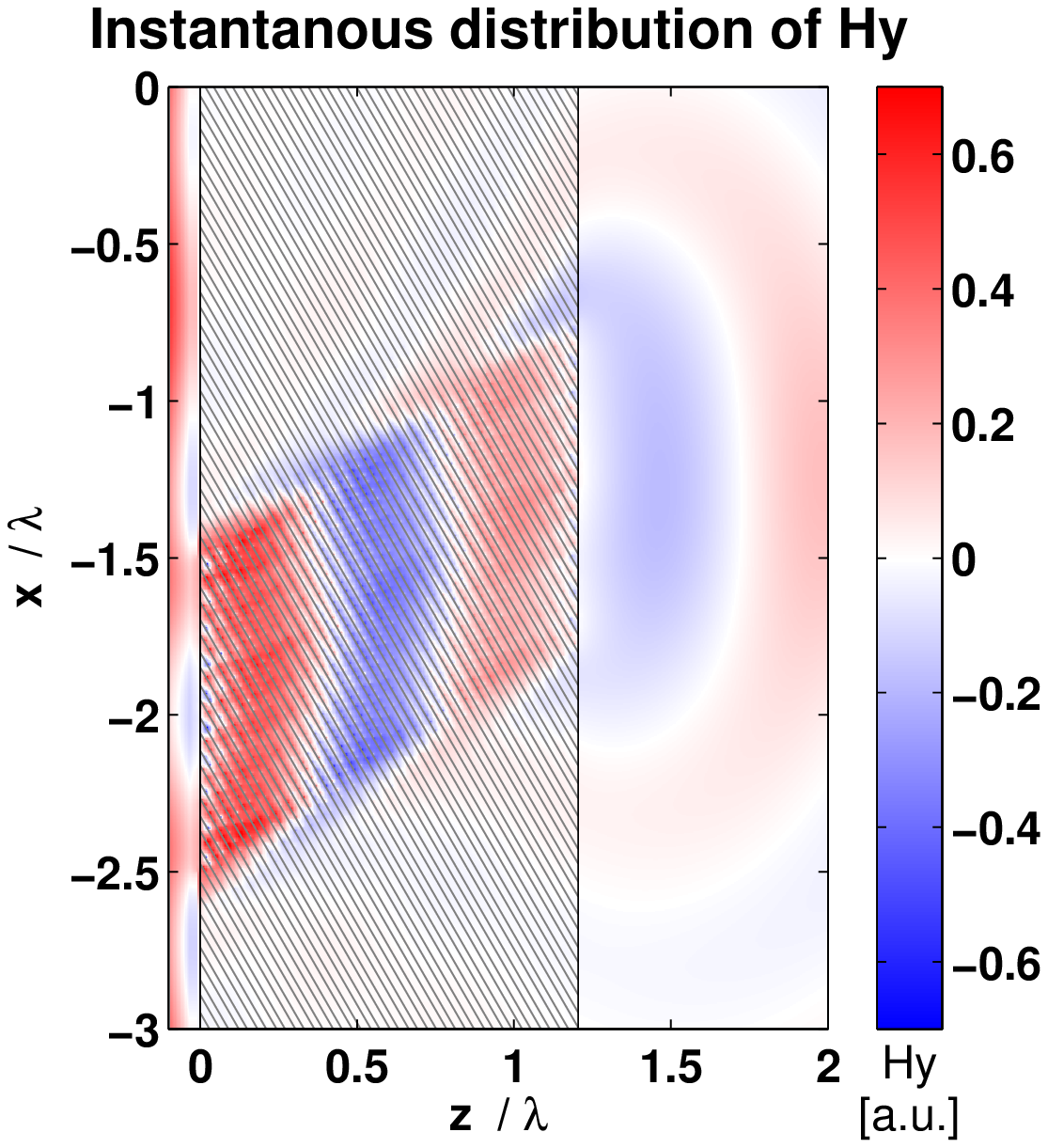}
\caption{Diffraction-free transmission of the subwavelength sized beam through the multilayer (FDTD simulations, with the incident CW beam limited by a subwavelength aperture in a perfect conductor): \textbf{a}.~time-averaged Poynting vector $S_z$ in a rectangular slab; \textbf{b,c}.~instantaneous magnetic field $H_y$ in a slab with layer boundaries oriented at $30^\circ$ towards the external boundaries for various aperture sizes.\label{fig.fdtd1}}
\end{figure} 

\begin{figure}
a.\includegraphics[height=5cm]{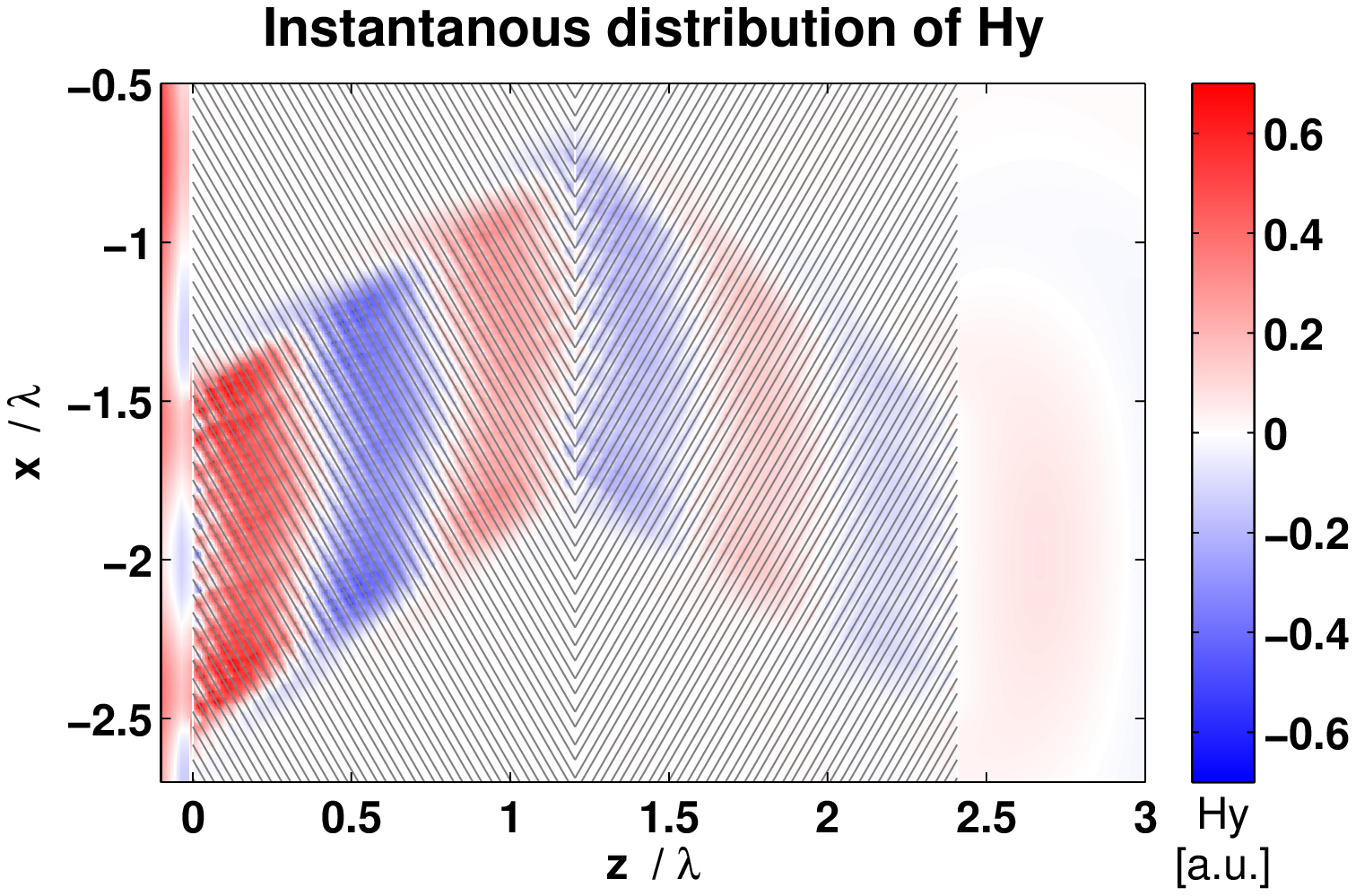}
b.\includegraphics[height=5cm]{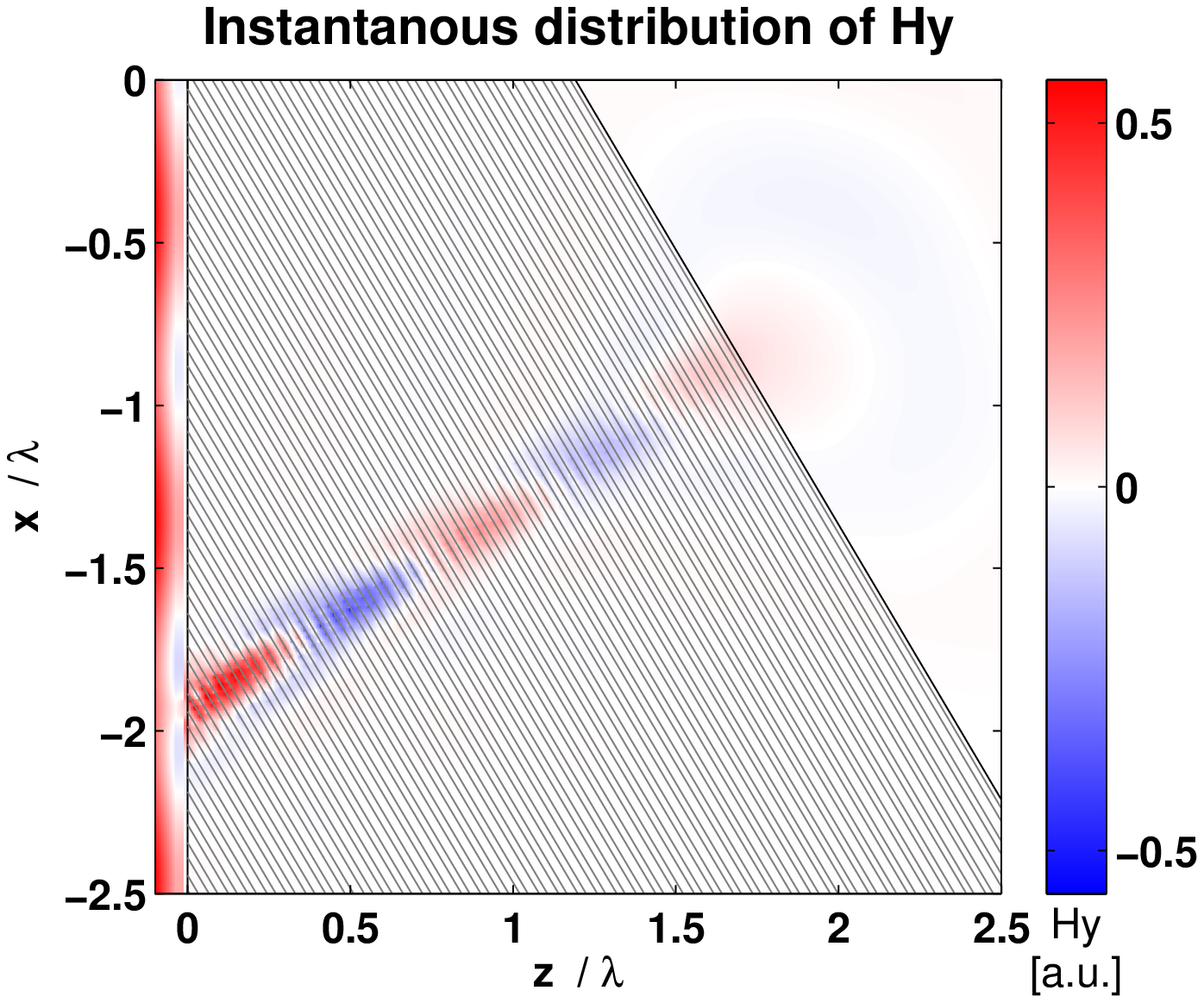}
\caption{Simple optical nano-elements internally made of the diffraction-free material \textbf{a}. double slab; \textbf{b}. prism;.
\label{fig.fdtd2}}
\end{figure}

\psection{Conclusions}
We have demonstrated a diffraction-free, low-loss material, which is impedance matched to air. The diffraction-free propagation  is verified by the analysis of FWHM of PSF in the function of the number of periods. The material  consists of silver and a dielectric with refractive index of $n=2.6$ and operates at $\lambda=422nm$. It has the effective imaginary part of refractive index smaller than that of silver by a factor of $50$. The reflections are in between $-10dB$ and $-30dB$, and the FWHM of PSF is at the order of $40nm$. Small reflections, small attenuation, and reduced Fabry Perot resonances make it a flexible metamaterial for arbitrarily shaped optical planar devices with sizes of the order of one wavelength, such as the elements of optical interconnects or cloaks. 

\ack
We acknowledge support from the Polish MNiI research projects \textit{N N202 033237} and \textit{N R15 0018 06}
 as well as the framework of COST actions \textit{MP0702} and \textit{MP0803}.

\end{paper} 
\end{document}